\documentclass[aps,prc,amssymb,amsmath,onecolumn,preprint]{revtex4-1}

\usepackage{graphicx}
\usepackage{float}
\usepackage{amsmath}
\usepackage{epstopdf}
\usepackage{amsfonts,amssymb}
\usepackage[utf8]{inputenc}
\usepackage{pstricks}
\usepackage{color}
\usepackage{hyperref}
\usepackage[toc]{appendix}

\usepackage{color, colortbl}
\definecolor{Gray}{gray}{0.9}
\usepackage{float}
\usepackage{tikz-feynman}
\usepackage{forest}

\usepackage{xcolor}
\usepackage{amssymb}
\usepackage{txfonts}
\usepackage{feynmp-auto}
\usepackage{braket}
\usepackage{tikz} 
\usepackage{simpler-wick}
\usetikzlibrary{shapes,arrows,positioning,automata,backgrounds,calc,er,patterns}
\usepackage[left=2.5cm,top=2.5cm,right=2.5cm,bottom=2.5cm]{geometry}
\tikzfeynmanset{compat=1.0.0}

\usepackage{verbatim}    
\usepackage{dcolumn}
\usepackage{bm}
\usepackage{epsfig}
\usepackage{braket}
\usepackage[normalem]{ulem} 
\usepackage{sidecap}
\usepackage[caption=false]{subfig}
\newcommand{\be}{\begin{equation}}
\newcommand{\ee}{\end{equation}}
\newcommand{\ben}{\begin{eqnarray}}
\newcommand{\een}{\end{eqnarray}}

\newcommand{\pslash}{\not{\hbox{\kern-2.3pt $p$}}}
\newcommand{\pdslash}{\not{\hbox{\kern-2pt $\partial$}}}

\begin{document}
\title{Production of the $ X(3872)$ state via the $B^0 \to K^{\ast 0}  X(3872)$ decay}

\author{Luciano M. Abreu}
\email{luciano.abreu@ufba.br}
\affiliation{ Instituto de F\'isica, Universidade Federal da Bahia,
Campus Universit\'ario de Ondina, 40170-115, Bahia, Brazil}

\begin{abstract}

In this work the production of the state $X(3872)$ is estimated via the reaction $B^0 \to K^{\ast 0}  X(3872)$ through triangle mechanisms described by the sequence $B^0 \to  D_s^{(*)+} (\to K^{\ast 0} D^{(*)+} ) \  D^{(*)-} \to K^{\ast 0} \ (  D^{(*)+} D^{(*)-} ) \to K^{\ast 0} X(3872) $. The molecular configuration $(D\bar D^* - c.c.  )$ of the $X(3872)$ is considered. By means of the effective Lagrangian approach, the branching ratio $\mathcal{B}(B^0 \to K^{\ast 0}  X(3872))$  is calculated as a function of the strength of the coupling of the charged components $(D^+\bar D^{*-} - c.c. )$ to the $X(3872)$ and compared with experimental data. Besides, employing the decay $B^0 \to K^{\ast 0}  \psi (2S)$ as a normalization channel, the ratio of branching fractions $R = \frac{\mathcal{B}( B^0 \to K^{\ast 0}  X(3872) )}{\mathcal{B}( B^0 \to K^{\ast 0} \psi (2S) )}\times \frac{\mathcal{B}( X(3872) \to J/\psi \pi^{+} \pi^{-} )}{\mathcal{B}( \psi (2S) \to J/\psi \pi^{+} \pi^{-} )} $ is also estimated. The findings provide another concrete example for the vital role of charged components in achieving a quantitatively correct description of the $X(3872)$.

\end{abstract}
\maketitle

%%%%%%%%%%%%%%%%%%%%%%%%%%%%%%%%%%%%%%%%%%%%%%%%%%%
%%%%%%%%%%%%%%%%%%%%%%%%%%%%%%%%%%%%%%%%%%%%%%%%%%%
%%%%%%%%%%%%%%%%%%%%%%%%%%%%%%%%%%%%%%%%%%%%%%%%%%%
\section{INTRODUCTION}
\label{sec-intr}
%%%%%%%%%%%%%%%%%%%%%%%%%%%%%%%%%%%%%%%%%%%%%%%%%%%
%%%%%%%%%%%%%%%%%%%%%%%%%%%%%%%%%%%%%%%%%%%%%%%%%%%
%%%%%%%%%%%%%%%%%%%%%%%%%%%%%%%%%%%%%%%%%%%%%%%%%%%

Numerous new hadrons have been discovered over the past few decades exhibiting exotic features that cannot be explained within the conventional quark model framework~\cite{ParticleDataGroup:2024cfk,Brambilla:2019esw,Chen:2022asf,Meng:2022ozq}. The underlying structure of these states has been interpreted in various ways, with proposed explanations ranging from weakly-bound hadronic molecules and compact multiquark systems to excited conventional hadrons, kinematical cusps, hybrid states, glueballs, and possible mixtures of these configurations. Despite extensive research, no universal consensus has emerged regarding their fundamental nature, making this field one of the most actively debated subjects in hadron physics. To discriminate among these different interpretations, theoretical and experimental studies have focused on key observables such as mass spectra, decay widths, final yields and production cross-sections~\cite{Brambilla:2019esw}.

The paradigmatic case in this context is the $X(3872)$ state (also denoted as $\chi_{c1}(3872)$), first observed in 2003 by Belle Collaboration and after confirmed by other experiments~\cite{Belle:2003nnu,ParticleDataGroup:2024cfk}, with quantum numbers $I^G(J^{PC}) = 0^+(1^{++})$. Its fundamental structure remains controversial, with the two leading candidates being: a loosely bound $(D\bar D^* + c.c.  )$  molecular state, and a compact $c\bar{c}q\bar{q}$, or an admixture of them~\cite{Brambilla:2019esw,Chen:2022asf,Meng:2022ozq,Swanson:2003tb,Tornqvist:2004qy,Dong:2009uf,Dong:2009yp,Cho:2013rpa,MartinezTorres:2014son,Abreu:2016qci,Guo:2017jvc,Ortega:2020qqm,Esposito:2020ywk,Braaten:2020iqw,Song:2023pdq,Abreu:2024mxc,Esposito:2025hlp}. 

Thus, the central aim that persists is to achieve a complete theoretical understanding of the $X(3872)$ state, encompassing its internal structure, spectroscopic properties, decay patterns, and its production in hadronic collisions. In particular, the decays of $B$ mesons into the $X(3872)$ have furnished an important testing ground for theoretical models describing its production mechanisms.
The comparison between its observed branching fractions and theoretical predictions provides a crucial test for the molecular state description. For instance, the $B \to X(3872)K$ reactions have been analyzed in Refs.~\cite{Braaten:2004fk,Braaten:2004ai}, considering the $B$ meson decaying into $\bar{D}^{*} D K$, followed by final-state rescattering of the charmed mesons to dynamically generate the $X(3872)$. However, the predicted branching ratio $\mathcal{B}(B^0 \to X(3872)K^0)$ has been one order of magnitude smaller than $\mathcal{B}(B^+ \to X(3872)K^+)$, in disagreement with the experimental measurement. On the other hand, while Ref.~\cite{Wang:2022xga} successfully described the ratio $\mathcal{B}(B^0 \to X(3872)K^0)/ \mathcal{B}(B^+ \to X(3872)K^+)$ by accounting for both charged and neutral meson loops, their approach could not simultaneously account for the absolute branching fractions. This discrepancy has been solved in Ref.~\cite{Wu:2023rrp} by proposing these production processes  $B^{+(0)} \to X(3872)K^{+(0)}$ via triangle diagram mechanisms with neutral and charged $\bar{D}^{*}D$ components, with the weak vertices modeled by means of $B \to \bar D^{(*)}$-transition form factors, and then demonstrating how the $(D\bar D^* - c.c.  )$  molecular picture provides a reasonable description of this $X(3872)$-production.

In addition to the production reactions mentioned above, the $X(3872)$ has also been detected in other $B_{(s)}$ decay modes. Other interesting channel is the $B^0 \to K^{\ast 0} X(3872)$, which has not been evaluated in Ref.~\cite{Wu:2023rrp}. Its branching ratio reported in the Review of Particle Physics is $(9 \pm 5 ) \times 10^{-5}$~\cite{ParticleDataGroup:2024cfk}. 
Moreover, using the $ B^0 \to K^{\ast 0}  \psi(2S)  $ decays as a normalization channel, the LHCb Collaboration reported 
in Fig.~2 of  Ref.~\cite{LHCb:2025vjj} the relative branching ratio between the $ \mathcal{B} (B^0 \to K^{\ast 0}  X(3872)) $ and $ B^0 \to K^{\ast 0}  \psi(2S)  $, which has been set to be 
\begin{eqnarray}
R = \frac{\mathcal{B}( B^0 \to K^{\ast 0}  X(3872) )}{\mathcal{B}( B^0 \to K^{\ast 0} \psi (2S) )}\times \frac{\mathcal{B}( X(3872) \to J/\psi \pi^{+} \pi^{-} )}{\mathcal{B}( \psi (2S) \to J/\psi \pi^{+} \pi^{-} )} = (1.95 \pm 0.75) \times 10^{-2}. 
\label{ratio}
\end{eqnarray}

In this scenario, the key point motivating the present work is the role of the charged meson components $(D^-D^{*+} - c.c.)$ of the $X(3872)$ in reactions as the $B^0 \to K^{\ast 0}  X(3872) $. 
This can be justified by the following reasoning (see Ref.~\cite{MartinezTorres:2014son} for a detailed discussion): for short-range hadronic transitions, the relevant observable is the wave function at the origin, which is directly related to the effective vertex coupling. Since the coupling to charged channel is close to the neutral one, a complete model must incorporate both. Thus, their inclusion prevents under-predicting decay widths to $J/\psi\rho$, $J/\psi\omega$, and $J/\psi\gamma$ when compared to data~\cite{Gamermann:2009uq,Aceti:2012cb}, and also the previously mentioned ratio $\mathcal{B}(B^0 \to X(3872)K^0)/ \mathcal{B}(B^+ \to X(3872)K^+)$~\cite{Wu:2023rrp}.

Thus, inspired by preceding analyses, the objective of this study is to provide another concrete case for the mandatory inclusion of the charged components, which is crucial for a quantitatively accurate understanding of the $X(3872)$, thereby complementing and extending the findings of Ref.~\cite{Wu:2023rrp}. Considering the $B^0 \to K^{\ast 0}  X(3872) $ reaction coming from a triangle mechanism, and assuming the $(D\bar D^* - c.c.  )$  molecular picture for the $ X(3872)$, this decay is described by the chain $B^0 \to  D_s^{(*)+} (\to K^{\ast 0} D^{(*)+} ) D^{(*)-} \to K^{\ast 0} (  D^{(*)+} D^{(*)-} ) \to K^{\ast 0} X(3872) $. Within an effective Lagrangian approach, the branching fraction $\mathcal{B}(B^0 \to K^{\ast 0} X(3872))$ is calculated considering its sensitivity to the coupling strength $g_{X D D^{*}}$ of the charged components $(D^+\bar{D}^{-} - \text{c.c.})$ to the $X(3872)$. The results are compared with available experimental data. Furthermore, we estimate the ratio $R$ given in Eq.~(\ref{ratio}).

The paper is organized as follows. Section~\ref{sec:Formalism} presents the formalism used to describe the $X(3872)$-production through $B^0 \to K^{\ast 0}  X(3872)$ decays. Section~\ref{Results} is devoted to show the numerical results and to analyze them. Concluding remarks are provided in Section~\ref{Conclusions}.

%%%%%%%%%%%%%%%%%%%%%%%%%%%%%%%%%%%%%%%%%%%%%%%%%%%%%%%%%%%%%%%%%%%%
%%%%%%%%%%%%%%%%%%%%%%%%%%%%%%%%%%%%%%%%%%%%%%%%%%%%%%%%%%%%%%%%%%%%
%%%%%%%%%%%%%%%%%%%%%%%%%%%%%%%%%%%%%%%%%%%%%%%%%%%%%%%%%%%%%%%%%%%%
\section{Formalism}
\label{sec:Formalism} 
%%%%%%%%%%%%%%%%%%%%%%%%%%%%%%%%%%%%%%%%%%%%%%%%%%%%%%%%%%%%%%%%%%%%
%%%%%%%%%%%%%%%%%%%%%%%%%%%%%%%%%%%%%%%%%%%%%%%%%%%%%%%%%%%%%%%%%%%%
%%%%%%%%%%%%%%%%%%%%%%%%%%%%%%%%%%%%%%%%%%%%%%%%%%%%%%%%%%%%%%%%%%%%

\begin{figure}[!htbp]
    \centering
%    \subfigure[]{
\begin{tikzpicture}[very thick,q0/.style={->,thick,yshift=5pt,shorten >=5pt,shorten <=5pt}]
\tikzfeynmanset{ every vertex = {dot} }
\begin{feynman}
    \vertex (a1){$\bar b$};
	\vertex[right=2.5cm of a1] (a2) ;
	\vertex[right=1.5cm of a2] (a3) {};
	\vertex[right=1.5cm of a3] (a4) {$\bar c$};
	\vertex[right=0.5cm of a4] (a41){};
	\vertex[right=0.5cm of a41] (a5){};
\vertex[below=1.cm of a1] (c1) {$d$};
\vertex[below=1.cm of a4] (c2){$d$};
\vertex[above=1.5cm of a3] (d1);
\vertex[above=1.0cm of a4] (d2){$c$};
\vertex[above=2.5cm of a4] (d3){$\bar s$};
  \diagram* {
  (a2) --  [fermion, edge label=  {}]  (a1), (a2) -- [boson, edge label= {$W^{+} $}](d1),(a4) -- [fermion, edge label= {}](a2), (d1) --  [fermion, edge label= {}](d2), (d3) --  [fermion, edge label= {}](d1),  (c1) --[fermion, edge label= {}]
   (c2), 
}; 
\end{feynman}
\end{tikzpicture}  
%  \label{quarklv}   
%}
\caption{ The weak decay relevant for the reaction to be evaluated. The $\bar b$ quark decays into a $\bar c$ quark by an external emission of a $W^{+}$ boson which decays into a $c\bar{s}$ pair to form a $D_s^{(\ast)+}$ meson. The $d$ quark is a spectator of the reaction and together with the $\bar c$ produces the $D^{(*)-}$. }
\label{DIAG1}
\end{figure}
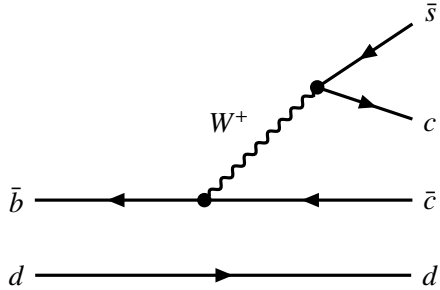

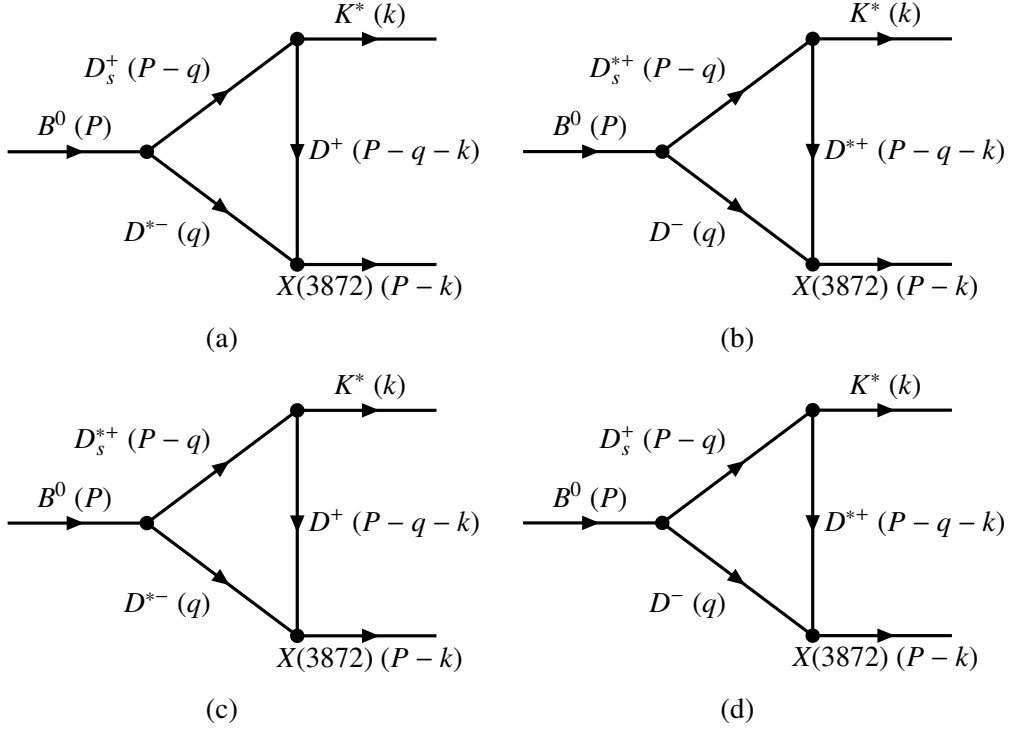
\begin{figure}[!htbp]
	\centering
\begin{tikzpicture}[very thick,q0/.style={->,semithick,yshift=5pt,shorten >=5pt,shorten <=5pt}]
\tikzfeynmanset{ every vertex = {dot} }
\begin{feynman}
    \vertex (a1){};
	\vertex[right=2.0cm of a1] (a2) ;
	\vertex[right=2.0cm of a2] (a3) {};
	\vertex[right=2.0cm of a3] (a4) {};
	\vertex[right=0.5cm of a4] (a41){};
	\vertex[right=3.0cm of a1] (a23) {} ;
\vertex[below=1.5cm of a3] (c1);
\vertex[below=1.5cm of a4] (c2){};
\vertex[below=2.5cm of a23] (c3){(a)};
\vertex[above=1.5cm of a3] (d1);
\vertex[above=1.5cm of a4] (d2){};
  \diagram* {
  (a1) --  [fermion, edge label=  {$B^0 \  (P)$}]  (a2), (a2) -- [fermion, edge label= {${D}_s^{+} \  (P-q)$}](d1), (d1) --  [fermion, edge label= {$K^\ast \ (k)$}](d2), (a2) --  [fermion, edge label'= {${D}^{\ast-} \ (q)$}](c1), (c1) --[fermion, edge label'= {$X(3872) \ (P-k)$}]
   (c2), (d1) --  [fermion, edge label= {${D}^{+}  \ (P-q-k)$}] (c1), 
}; 
\end{feynman}
\end{tikzpicture}
\begin{tikzpicture}[very thick,q0/.style={->,semithick,yshift=5pt,shorten >=5pt,shorten <=5pt}]
\tikzfeynmanset{ every vertex = {dot} }
\begin{feynman}
    \vertex (a1){};
	\vertex[right=2.0cm of a1] (a2) ;
	\vertex[right=2.0cm of a2] (a3) {};
	\vertex[right=2.0cm of a3] (a4) {};
	\vertex[right=0.5cm of a4] (a41){};
	\vertex[right=3.0cm of a1] (a23) {} ;
\vertex[below=1.5cm of a3] (c1);
\vertex[below=1.5cm of a4] (c2){};
\vertex[below=2.5cm of a23] (c3){(b)};
\vertex[above=1.5cm of a3] (d1);
\vertex[above=1.5cm of a4] (d2){};
  \diagram* {
  (a1) --  [fermion, edge label=  {$B^0 \  (P)$}]  (a2), (a2) -- [fermion, edge label= {${D}_s^{\ast +} \  (P-q)$}](d1), (d1) --  [fermion, edge label= {$K^\ast \ (k)$}](d2), (a2) --  [fermion, edge label'= {${D}^{-} \ (q)$}](c1), (c1) --[fermion, edge label'= {$X(3872) \ (P-k)$}]
   (c2), (d1) --  [fermion, edge label= {${D}^{\ast +}  \ (P-q-k)$}] (c1), 
}; 
\end{feynman}
\end{tikzpicture}
\begin{tikzpicture}[very thick,q0/.style={->,semithick,yshift=5pt,shorten >=5pt,shorten <=5pt}]
\tikzfeynmanset{ every vertex = {dot} }
\begin{feynman}
    \vertex (a1){};
	\vertex[right=2.0cm of a1] (a2) ;
	\vertex[right=2.0cm of a2] (a3) {};
	\vertex[right=2.0cm of a3] (a4) {};
	\vertex[right=0.5cm of a4] (a41){};
	\vertex[right=3.0cm of a1] (a23) {} ;
\vertex[below=1.5cm of a3] (c1);
\vertex[below=1.5cm of a4] (c2){};
\vertex[below=2.5cm of a23] (c3){(c)};
\vertex[above=1.5cm of a3] (d1);
\vertex[above=1.5cm of a4] (d2){};
  \diagram* {
  (a1) --  [fermion, edge label=  {$B^0 \  (P)$}]  (a2), (a2) -- [fermion, edge label= {${D}_s^{\ast +} \  (P-q)$}](d1), (d1) --  [fermion, edge label= {$K^\ast \ (k)$}](d2), (a2) --  [fermion, edge label'= {${D}^{\ast-} \ (q)$}](c1), (c1) --[fermion, edge label'= {$X(3872) \ (P-k)$}]
   (c2), (d1) --  [fermion, edge label= {${D}^{+}  \ (P-q-k)$}] (c1), 
}; 
\end{feynman}
\end{tikzpicture}
\begin{tikzpicture}[very thick,q0/.style={->,semithick,yshift=5pt,shorten >=5pt,shorten <=5pt}]
\tikzfeynmanset{ every vertex = {dot} }
\begin{feynman}
    \vertex (a1){};
	\vertex[right=2.0cm of a1] (a2) ;
	\vertex[right=2.0cm of a2] (a3) {};
	\vertex[right=2.0cm of a3] (a4) {};
	\vertex[right=0.5cm of a4] (a41){};
	\vertex[right=3.0cm of a1] (a23) {} ;
\vertex[below=1.5cm of a3] (c1);
\vertex[below=1.5cm of a4] (c2){};
\vertex[below=2.5cm of a23] (c3){(d)};
\vertex[above=1.5cm of a3] (d1);
\vertex[above=1.5cm of a4] (d2){};
  \diagram* {
  (a1) --  [fermion, edge label=  {$B^0 \  (P)$}]  (a2), (a2) -- [fermion, edge label= {${D}_s^{ +} \  (P-q)$}](d1), (d1) --  [fermion, edge label= {$K^\ast \ (k)$}](d2), (a2) --  [fermion, edge label'= {${D}^{-} \ (q)$}](c1), (c1) --[fermion, edge label'= {$X(3872) \ (P-k)$}]
   (c2), (d1) --  [fermion, edge label= {${D}^{\ast+}  \ (P-q-k)$}] (c1), 
}; 
\end{feynman}
\end{tikzpicture}

\captionsetup{width=\linewidth, format=hang}
\caption{Triangle Feynman diagrams for the decay  $B^0 \to K^{\ast 0}  X(3872)$  via the following charmed meson loops: (a) $B^0 \to  D_s^{+} (\to K^{\ast 0} D^{+} ) D^{*-} \to K^{\ast 0} (  D^+ D^{*-} ) \to K^{\ast 0} X(3872) $;  (b) $B^0 \to  D_s^{\ast +} (\to K^{\ast 0} D^{\ast +} ) D^{-} \to K^{\ast 0} (  D^{\ast +} D^{-} ) \to K^{\ast 0} X(3872) $;  (c) $B^0 \to  D_s^{\ast +} (\to K^{\ast 0} D^{+} ) D^{*-} \to K^{\ast 0} (  D^+ D^{*-} ) \to K^{\ast 0} X(3872) $; (d) $B^0 \to  D_s^{ +} (\to K^{\ast 0} D^{\ast +} ) D^{-} \to K^{\ast 0} (  D^{* +} D^{-} ) \to K^{\ast 0} X(3872) $. The respective momenta of the particles are in parentheses. }
\label{DIAG2}
\end{figure}

The starting point of this study is the assumption that the state $X(3872)$, characterized by the quantum numbers $I^G(J^{PC})=0^+(1^{++})$, is mainly constituted as a quantum superposition of the following $S$-wave molecular components $(D^{*0} \bar{D}^0-c.c.)$ and  $( D^{*+}D^- -c.c.)$~\cite{Dong:2009uf,Dong:2009yp,MartinezTorres:2014son,Sakai:2020ucu,Song:2023pdq,Wu:2023rrp}.
Accordingly, for the $X(3872)$-production through $B^0 \to K^{\ast 0}  X(3872)$ decays, the relevant weak decay to be considered is shown in Fig.~\ref{DIAG1}. The $\bar b$ quark decays into a $\bar c$ quark by an external emission of a $W^{+}$ boson, which decays into a $c\bar{s}$ pair to form a $D_s^{(\ast)+}$ meson. The $d$ quark from the $B^0$ is a spectator of the reaction and will constitute the $D^{(*)-}$. At the hadron level, the final state is produced from the initial weak vertex above mentioned, through the triangle mechanisms described by the chains depicted in Fig.~\ref{DIAG2}.

%The formalism used to study the production of the state $X(3872)$ through $B^0 \to K^{\ast 0}  X(3872)$ decays is presented here, under the assumption  that it is mainly constituted as a quantum superposition of the following $S$-wave molecular components~\cite{Dong:2009uf,Dong:2009yp,MartinezTorres:2014son,Sakai:2020ucu,Song:2023pdq,Wu:2023rrp},
%\begin{align}
%\ket{X(3872)} &= 
%\frac{Z_{N}^{1/2}}{\sqrt{2}} \left( \ket{D^0 \bar{D}^{*0}} + \ket{D^{*0} \bar{D}^0} \right) 
%+ \frac{Z_{C}^{1/2}}{\sqrt{2}} \left( \ket{D^+ D^{*-}} + \ket{D^- D^{*+}} \right),
%\label{eq:x_composition}
%\end{align}
%where the factors $Z_N, Z_C$ represent the probability amplitudes for the neutral and charged configurations, respectively. Accordingly, Fig.~\ref{DIAG1} shows the weak decay relevant for the reaction to be evaluated. The $\bar b$ quark decays into a $\bar c$ quark by an external emission of a $W^{+}$ boson which decays into a $c\bar{s}$ pair to form a $D_s^{(\ast)+}$ meson. The $d$ quark is a spectator of the reaction and produce the $D^{(*)-}$. At the hadron level, the final state is produced from the initial weak vertex above mentioned, through a triangle mechanism described by the chain $B^0 \to  D_s^+ (\to K^{\ast 0} D^+ ) D^{*-} \to K^{\ast 0} (  D^+ D^{*-} ) \to K^{\ast 0} X(3872) $. This reaction is depicted in Fig.~\ref{DIAG2}.

%%%%%%%%%%%%%%%%%%%%%%%%%%%%%%%%%%%%%%%%%%%%%%%%%%%%%%%%%%%%%%%%%%%%
%%%%%%%%%%%%%%%%%%%%%%%%%%%%%%%%%%%%%%%%%%%%%%%%%%%%%%%%%%%%%%%%%%%%
%%%%%%%%%%%%%%%%%%%%%%%%%%%%%%%%%%%%%%%%%%%%%%%%%%%%%%%%%%%%%%%%%%%%
%%%%%%%%%%%%%%%%%%%%%%%%%%%%%%%%%%%%%%%%%%%%%%%%%%%%%%%%%%%%%%%%%%%%

%%%%%%%%%%%%%%%%%%%%%%%%%%%%%%%%%%%%%%%%%%%%%%%%%%%%%%%%%%%%%%%%%%%%
%%%%%%%%%%%%%%%%%%%%%%%%%%%%%%%%%%%%%%%%%%%%%%%%%%%%%%%%%%%%%%%%%%%%
%%%%%%%%%%%%%%%%%%%%%%%%%%%%%%%%%%%%%%%%%%%%%%%%%%%%%%%%%%%%%%%%%%%%
%%%%%%%%%%%%%%%%%%%%%%%%%%%%%%%%%%%%%%%%%%%%%%%%%%%%%%%%%%%%%%%%%%%%

The calculation of the amplitudes represented by the Feynman diagrams of Fig.~\ref{DIAG2} will be performed within the framework of the effective Lagrangian approach, as discussed below.

%%%%%%%%%%%%%%%%%%%%%%%%%%%%%%%%%%%%%%%%%%%%%%%%%%%%%%%%%%%%%%%%%%%%
%%%%%%%%%%%%%%%%%%%%%%%%%%%%%%%%%%%%%%%%%%%%%%%%%%%%%%%%%%%%%%%%%%%%
\subsection{The weak interaction vertices $B^0 \to  D_s^{(\ast) +} D^{(*) -}$} \label{subsec-weak}
%%%%%%%%%%%%%%%%%%%%%%%%%%%%%%%%%%%%%%%%%%%%%%%%%%%%%%%%%%%%%%%%%%%%
%%%%%%%%%%%%%%%%%%%%%%%%%%%%%%%%%%%%%%%%%%%%%%%%%%%%%%%%%%%%%%%%%%%%

To model the decay, the initial weak production and the subsequent hadronic final-state interactions will be distinguished, with the primary weak decays $B^0 \to D_s^{(*)+} D^{(*)-}$ being treated within a effectively framework. A detailed justification of this methodology, which draws inspiration from the factorization approach~\cite{Ali:1998eb,Wu:2023rrp}, can be found in Appendix~\ref{sec:appa}.
The adopted effective approach prioritizes the essential weak process dynamics, providing a simplified and tractable framework for an initial study of the rescattering effects that produce the final state of interest.

On more concrete grounds, the production amplitudes for the decay modes $B^0 (P) \to D_s^{(*)+}(p_1) D^{(*)-}(p_2)$ are parameterized as follows:
\begin{eqnarray}
 \mathcal{M}( B^0 \to  D_s^+  D^{*-})  & = &  g_{ B D_s \bar D^{\ast} }  (P+p_1)_{\mu} \epsilon_{D^{*-}}^\mu (p_2)  , \nonumber \\
\mathcal{M}( B^0 \to  D_s^{\ast +}  D^{-})  & = &  g_{ B D_s^{\ast} \bar D}  (P+p_2)_{\mu} \epsilon_{D_s^{\ast+}}^\mu (p_1)  , \nonumber \\
\mathcal{M}( B^0 \to  D_s^{\ast +}  D^{\ast-})  & = & g_{ B D_s^{\ast} \bar D^{\ast}}  \varepsilon _{\mu \nu \rho \sigma }  p_{1}^{\mu}p_{2}^{\rho} \epsilon_{D_s^{\ast+}}^{\nu} (p_1)  \epsilon_{D^{\ast-}}^{ \sigma} (p_2), 
\nonumber \\
\mathcal{M}( B^0 \to  D_s^+  D^{-})  & = &  g_{ B D_s \bar D }  , 
  \label{amplBDDS}
\end{eqnarray}
where $g_{ B D_s^{(\ast)} \bar D^{(\ast)} }$ are the effective coupling constants to be determined, and $\epsilon ^\alpha(p_i)$ are the polarization vectors.
So, from the expression for the decay rate 
\begin{eqnarray} \Gamma( B^0 \to  D_s^{(\ast)+}  D^{(*)-} ) = \frac{1}{8\pi}\frac{|\vec{p_1}|}{ m^2_{B}} \sum_{Pol} |  \mathcal{M}( B^0 \to  D_s^{(\ast)+}  D^{(*)-} ) |^2 ,
  \label{decayrate1}
\end{eqnarray}
where  $|\vec{p}| = \lambda ^{1/2} \left(m^2_{B},m^2_{D_s^{(\ast)} },m^2_{\bar D^{(\ast)}} \right) / (2 m_{B})$ is the magnitude of the three-momentum of the $D_s^{(\ast)+}$ meson in the rest frame of $B^0$; $ \lambda (a,b,c) = a^2 + b^2 + c^2 - 2 a b - 2 a c - 2 b c $ is the K\"allen function; and $\sum_{Pol}$ denotes the sum over the polarizations of the final state. Thus, substituting the amplitudes given in Eq.~(\ref{amplBDDS}) into Eq.~(\ref{decayrate1}), one can determine the coupling constants from the relation 
\begin{eqnarray}
 \frac{  g_{ B D_s^{(\ast)} \bar D^{(\ast)} }^2 }{\Gamma_ {B}} = \frac{2 \pi A}{|\vec{p_1}|^3} \times  
 \mathcal{B}( B^0 \to   D_s^{(\ast)+}  D^{(*)-} ), 
  \label{couplBDSD}
\end{eqnarray}
where $A( B^0 \to   D_s^{+}  D^{*-} ) = m_ {D^{*-}}^2$, $A( B^0 \to   D_s^{* +}  D^{-} )=  m_ {D_s^{*+}}^2$, $A( B^0 \to   D_s^{* +}  D^{* -} ) = 2 $, and  $A( B^0 \to   D_s^{ +}  D^{ -} ) = 4 m_B^2 |\vec{p_1}|^2$;
$ \mathcal{B} = \Gamma / \Gamma_ {B} $ is the branching ratio of the corresponding reaction ($\Gamma_ {B}$ is the total decay witdh). The experimental data of the relevant  branching ratios are  available in Ref.~\cite{ParticleDataGroup:2024cfk}, and summarized in Table~\ref{tableBR}. The coupling constants, obtained by fitting them to the central values of the branching ratios for the corresponding reactions available in Ref.~\cite{ParticleDataGroup:2024cfk}, are displayed in Table~\ref{tablecouplings}. 
%\begin{eqnarray}
% A( B^0 \to   D_s^{+}  D^{*-} )=  \frac{2 \pi m_ {D^{*}}^2}{|\vec{p_1}|^3} , \ A( B^0 \to   D_s^{* +}  D^{-} )=  \frac{2 \pi m_ {D_s^{*}}^2}{|\vec{p_1}|^3}, \  A( B^0 \to   D_s^{* +}  D^{-} )=  \frac{4 \pi }{|\vec{p_1}|^3} . 
%  \label{factorA}
%\end{eqnarray}

%%%%%%%%%%%%%%%%%%%%%%%%%%%%%%%%%%%%%%%%%%%%%%%%%%%%%%%%%%%%%%%%%%%%
\begin{table}[htbp!]
\centering
\caption{The experimental data available in Ref.~\cite{ParticleDataGroup:2024cfk} of the branching ratios for the reactions of interest. }
\begin{tabular}{ c c  }
 \hline\hline
       Decay mode  & $\mathcal{B} \ \  (10^{-3})$  \\
 \hline      
        $B^0 \to   D_s^{+}  D^{*-}$  &  $8.2 \pm 0.8$ \\
%        \hline
        $B^0 \to   D_s^{* +}  D^{-} $ &  $7.4 \pm 1.6$ \\
%        \hline
        $B^0 \to   D_s^{* +}  D^{* -}$  & $17.7 \pm 1.4$ \\
        $B^0 \to   D_s^{ +}  D^{-} $ &  $8.1 \pm 0.6$  \\
        $B^0 \to K^{\ast 0}  X(3872)$   & $0.09 \pm 0.05 $  \\
        $B^0 \to K^{\ast 0} \psi (2S) $ & $0.59 \pm 0.04 $  \\
        $X(3872) \to J/\psi \pi^{+} \pi^{-}$ & $43 \pm 14 $  \\
        $\psi (2S) \to J/\psi \pi^{+} \pi^{-}$ & $346.9 \pm 34 $  \\
\hline\hline
    \end{tabular}
 \label{tableBR}
\end{table}
%%%%%%%%%%%%%%%%%%%%%%%%%%%%%%%%%%%%%%%%%%%%%%%%%%%%%%%%%%%%%%%%%%%%

%%%%%%%%%%%%%%%%%%%%%%%%%%%%%%%%%%%%%%%%%%%%%%%%%%%%%%%%%%%%%%%%%%%%
\begin{table}[htbp!]
\centering
\caption{Coupling constants used for the vertices depicted in Fig.~\ref{DIAG2}. For the constants $g_{ B D_s^{(\ast)} \bar D^{(\ast)} }$, their squared values are normalized by the factor $\frac{ 1 }{\sqrt{\Gamma_ {B}}}$ (see Eq.~(\ref{couplBDSD})). }
 \begin{tabular}{ c c  }
 \hline\hline
      Coupling constant  & Value   \\
 \hline      
%        $ \frac{  g_{ B D_s \bar D^{\ast} }^2 }{\Gamma_ {B}} $  & $3.98 \times 10^{-5} \ \mathrm{MeV}^{-1}$ \\
        $ \frac{  g_{ B D_s \bar D^{\ast} } }{\sqrt{\Gamma_ {B}} } $  & $(6.31 \pm 0.31) \times 10^{-3} \ \mathrm{MeV}^{-1/2}$ \\
%        \hline
%        $\frac{  g_{ B D_s^{\ast} \bar D }^2 }{\Gamma_ {B}}$ & $3.95 \times 10^{-5} \ \mathrm{MeV}^{-1}$ \\
        $\frac{  g_{ B D_s^{\ast} \bar D } }{\sqrt{\Gamma_ {B}}}$ & $(6.29 \pm 0.68) \times 10^{-3} \ \mathrm{MeV}^{-1/2}$ \\%        \hline
%        $\frac{  g_{ B D_s^{\ast} \bar D^{\ast} }^2 }{\Gamma_ {B}}$  & $4.93 \ \times 10^{-11} \ \mathrm{MeV}^{-3}$ \\
        $\frac{  g_{ B D_s^{\ast} \bar D^{\ast} } }{\sqrt{\Gamma_ {B}}}$  & $( 7.02 \pm 0.28) \times 10^{-6} \ \mathrm{MeV}^{-3/2}$ \\
       $\frac{  g_{ B D_s \bar D } }{\sqrt{\Gamma_ {B}}}$  & $(55.95 \pm 2.07) \  \mathrm{MeV}^{1/2}$ \\
$g_{ D_s D K^{*} }$ & $4.0 \pm 0.4 $ \\
$g_{ D_s^{*} D K^{*}}$ & $(6.0 \pm 0.6)\times 10^{-3} \ \mathrm{MeV}^{-1}$ \\
$g_{ D_s^{*} D^{*} K^{*}}$  & $5.0\pm 0.5$ \\
$g_{ D_s D^{*} K^{*}}$ & $(6.0 \pm 0.6)\times 10^{-3} \ \mathrm{MeV}^{-1}$ \\
$ g_{X D^{*} \bar D}$   & $2000-7000 \ \mathrm{MeV}$ \\
 \hline\hline
\end{tabular} 
 \label{tablecouplings}
\end{table}
%%%%%%%%%%%%%%%%%%%%%%%%%%%%%%%%%%%%%%%%%%%%%%%%%%%%%%%%%%%%%%%%%%%%

%%%%%%%%%%%%%%%%%%%%%%%%%%%%%%%%%%%%%%%%%%%%%%%%%%%%%%%%%%%%%%%%%%%%
%%%%%%%%%%%%%%%%%%%%%%%%%%%%%%%%%%%%%%%%%%%%%%%%%%%%%%%%%%%%%%%%%%%%
\subsection{The interaction vertices $D_s^{(\ast) +} D^{(*) +} K^{* 0}$ }
\label{subsec-DsDKS}
%%%%%%%%%%%%%%%%%%%%%%%%%%%%%%%%%%%%%%%%%%%%%%%%%%%%%%%%%%%%%%%%%%%%
%%%%%%%%%%%%%%%%%%%%%%%%%%%%%%%%%%%%%%%%%%%%%%%%%%%%%%%%%%%%%%%%%%%%

To describe the interactions between charmed mesons and the vector kaon, the effective Lagrangians are~\cite{Azevedo:2004sv,Wang:2007zm,Abreu:2017cof,Janbazi:2018bgv,Aliev:2021cjt,CerqueiraJr:2021afn}:
\begin{eqnarray}
    \mathcal{L}_{D_s D K^{*}} & = & -i g_{ D_s D K^{*} } K^{\ast \mu} \left(  D \partial_\mu \bar D_s -  \partial_\mu  D \bar D_s 
    \right)  + h.c. , \nonumber \\
    \mathcal{L}_{D_s^{*} D K^{*}} & = & - g_{ D_s^{*} D K^{*}}  \varepsilon ^{\mu \nu \rho \sigma } \ D \ \partial_\mu K_{ \nu }^{\ast } \partial_\rho \bar D_{s\sigma}^{\ast } + h.c. , \nonumber \\
    \mathcal{L}_{D_s^{*} D^{*} K^{*}} & = & - i g_{ D_s^{*} D^{*} K^{*}}   \left[ K^{\ast \mu} \left(  D^{* \nu} \partial_\mu \bar D_{s \nu}^* - \partial_\mu D^{* \nu}  \bar D_{s \nu}^* \right) 
    + \left(  K^{* \nu} \partial_\mu D_{ \nu}^* -   \partial_\mu K^{* \nu} D_{ \nu}^* \right) \bar D_{s }^{\ast \mu} \right. \nonumber \\ 
   & & \left. + D^{\ast \mu} \left( \partial_\mu K_{ \nu}^*  \bar D_s^{* \nu}  - K_{ \nu}^* \partial_\mu  \bar D_s^{* \nu} \right) \right]  + h.c.
 \nonumber \\
 \mathcal{L}_{D_s D^{*} K^{*}}  & \textcolor{blue}{  = } &  - g_{ D_s D^{*} K^{*}}  \varepsilon ^{\mu \nu \rho \sigma } \ D_s \ \partial_\mu K_{ \nu }^{\ast } \partial_\rho \bar D_{\sigma}^{\ast } + h.c. , 
\label{LKD_sD}
\end{eqnarray}
with $g_{ D_s^{(*)} D^{(*)} K^{*} }$ being the coupling constants to be estimated. They have been calculated using different approaches. For example, the coupling $g_{ D_s D K^{*} }$ has been computed to be $g_{ D_s D K^{*} }=5.0$ in the $SU(4)$-flavor symmetry context~\cite{Azevedo:2003qh,Azevedo:2004sv}; 
%$1.61 \pm 0.32$ in the light-cone Sum Rules (LCSR) approach~\cite{Wang:2007zm}; 
$3.42\pm 0.44$~\cite{Khosravi:2015pfa}, and $3.09 \pm 0.50$~\cite{Janbazi:2017mpb} 
%and $2.29\substack{+0.65 \\ -0.41}$~\cite{CerqueiraJr:2021afn}
in distinct QCD sum rules calculations. In the case of  $g_{ D_s^{*} D K^{*}}$, it is determined to be $6.9 \ \mathrm{GeV}^{-1}$ via the $SU(4)$-flavor symmetry~\cite{Azevedo:2003qh,Azevedo:2004sv}; and $(3.74\pm 1.38)\ \mathrm{GeV}^{-1}$~\cite{Azizi:2010jj} with the QCD sum rules.
Similarly, $g_{ D_s D^{*} K^{*}}$ is $6.9 \ \mathrm{GeV}^{-1}$ in the $SU(4)$-flavor symmetry framework~\cite{Azevedo:2003qh,Azevedo:2004sv}; $(4.71\pm 0.39) \ \mathrm{GeV}^{-1}$~\cite{Khosravi:2015pfa}, and $(4.10 \pm 0.67) \ \mathrm{GeV}^{-1}$~\cite{Janbazi:2017mpb} 
in QCD sum rules calculations. 
For the coupling $g_{ D_s^{*} D^{*} K^{*}}$, it is estimated to be 5.0 with the $SU(4)$-flavor symmetry~\cite{Azevedo:2003qh,Azevedo:2004sv}; $4.77\pm 0.63$~\cite{Janbazi:2018bgv} and $5.20 \pm 0.70$~\cite{Khosravi:2015pfa} in the context of the QCD sum rules. In light of these discrepancies, a prudent strategy is adopted, whereby the chosen coupling parameters are selected to reside within the ranges established by prior mentioned studies; they are given in Table~\ref{tablecouplings}. 
%The uncertainties will be taken into account using a range for other parameters, as described further below. 

%%%%%%%%%%%%%%%%%%%%%%%%%%%%%%%%%%%%%%%%%%%%%%%%%%%%%%%%%%%%%%%%%%%%
%%%%%%%%%%%%%%%%%%%%%%%%%%%%%%%%%%%%%%%%%%%%%%%%%%%%%%%%%%%%%%%%%%%%
\subsection{The interaction vertex $X D^{*\pm} \bar D^{\mp} $}
\label{subsec-XDDStar}
%%%%%%%%%%%%%%%%%%%%%%%%%%%%%%%%%%%%%%%%%%%%%%%%%%%%%%%%%%%%%%%%%%%%
%%%%%%%%%%%%%%%%%%%%%%%%%%%%%%%%%%%%%%%%%%%%%%%%%%%%%%%%%%%%%%%%%%%%

The last ingredient necessary for the calculation of the amplitudes shown in Fig.~\ref{DIAG2} is the coupling of the $X(3872)$ state to the
hadron components $(D^{*} \bar D - c.c.)$. As previously mentioned, the hypothesis adopted is that the wave function of the $X(3872)$ around the origin (which is proportional to the couplings)  is very close to the combination of the components
$(D^{*0} \bar{D}^0-c.c.)$ and  $( D^{*+}D^- -c.c.)$, i.e.~\cite{Dong:2009uf,Dong:2009yp,MartinezTorres:2014son,Sakai:2020ucu,Song:2023pdq,Wu:2023rrp,Liu:2024ziu}
\begin{align}
\ket{X(3872)} &= 
\frac{1}{2} \left[ \ket{D^{*0} \bar{D}^0} + \ket{ D^{*+}D^-} - \ket{D^0 \bar{D}^{*0}} -  \ket{D^+ D^{*-}}  \right]. 
\label{eq:x_composition}
\end{align}
The phase convention for the doublets follows Ref.~\cite{Sakai:2020ucu}:  $(D^+,-D^0)$, $(\bar D^0,D^-)$ (and similarly for $D^*$);  and  $C D^+ = D^-$,  $C D^{* +} = -  D^{*-}$, were $C$ is the $C$-parity operator.

Therefore, the effective Lagrangian describing the interactions between the 
$X(3872)$ and its constituents are
\begin{eqnarray}
    \mathcal{L}_{X D^{*} \bar D^{}} & = &  g_{X D^{*} \bar D} X^{ \mu}  D^{*}_\mu \bar D  - h.c. , 
\label{LXDD}
\end{eqnarray}
with $ g_{X D^{*} \bar D}$ being the coupling constant to be estimated. As shown in Fig.~\ref{DIAG2}, the structure needed involves only the charged hadron components. In this sense, from different versions of a unitarized coupled channel approach, where $X(3872)$ is generated from the
dynamics of the respective hadron components, $ g_{X D^{*} \bar D}$ is estimated to be $2982 \ \mathrm{MeV}$~\cite{Gamermann:2009fv},  $3238 \ \mathrm{MeV}$~\cite{Gamermann:2009uq}, $3638 \ \mathrm{MeV}$~\cite{Aceti:2012cb}, $3000 \ \mathrm{MeV}$~\cite{Sakai:2020ucu}, $3390 \ \mathrm{MeV}$~\cite{Wu:2023rrp}. From the compositeness condition, it has been estimated to be  $ 9980 \sqrt{Z_c} \ \mathrm{MeV}$~\cite{Dong:2009yp}, with $Z_c = 0.033$ representing the probability amplitude for the neutral configuration.
In Ref.~\cite{Liu:2024ziu}, the value obtained by the contact-range effective field theory approach and Weinberg’s compositeness theorem has been $ 8290 \ \mathrm{MeV}$ and $ 9160 \ \mathrm{MeV}$, respectively. 
Due to the considerable fluctuations in the cited results, again  a conservative methodology is employed here, by considering  $ g_{X D^{*} \bar D}$  within the range of values reported above and shown in Table~\ref{tablecouplings}.

%%%%%%%%%%%%%%%%%%%%%%%%%%%%%%%%%%%%%%%%%%%%%%%%%%%%%%%%%%%%%%%%%%%%
%%%%%%%%%%%%%%%%%%%%%%%%%%%%%%%%%%%%%%%%%%%%%%%%%%%%%%%%%%%%%%%%%%%%
\subsection{The amplitudes and decay rate of the $B^0 \to K^{\ast 0}  X(3872)$ reaction }
\label{subsec-Ampl}
%%%%%%%%%%%%%%%%%%%%%%%%%%%%%%%%%%%%%%%%%%%%%%%%%%%%%%%%%%%%%%%%%%%%
%%%%%%%%%%%%%%%%%%%%%%%%%%%%%%%%%%%%%%%%%%%%%%%%%%%%%%%%%%%%%%%%%%%%

Thus, making use of the vertices discussed above, the amplitude of the $B^0 \to K^{\ast 0}  X(3872)$ decay can be written as
\begin{align}
% \mathcal{M}(B^0 \to K^{\ast 0}  X(3872)) & = \mathcal{M}_a +\mathcal{M}_b +\mathcal{M}_c
 \mathcal{M} & = \mathcal{M}_a +\mathcal{M}_b +\mathcal{M}_c +\mathcal{M}_d ,
\label{ampl1}
\end{align}
where $\mathcal{M}_i$ are are the contributions represented by the Feynman diagrams displayed in Fig.~\ref{DIAG2}; explicitly, 
\begin{align}
    \mathcal{M}_a &= -i \ g_{ B D_s \bar D^{\ast}}  g_{ D_s D K^{*} }  \left( \frac{g_{X D^{*} \bar D}}{2}\right) \epsilon_{K^*\ \mu}(k) \ \epsilon_{X}^\beta (P-k)  \nonumber \\ 
    & \times
    \int \frac{d^4q}{(2\pi)^4} 
\frac{(2P-q )_\nu [2(P-q)-k ]_\beta}{\left[ q^2- m_{\bar D^*}^2 \right] \left[ (P-q)^2-m_{D_s}^2 \right] \left[ (P-q-k)^2- m_{D}^2 \right] }
%\nonumber \\ 
%    & \times 
\left( -g^{\mu \nu } + \frac{q^\mu q^\nu}{m_{\bar D^*}^2}\right) , 
\nonumber \\     
\mathcal{M}_b &= - i \ g_{ B D_s^{\ast} \bar D}  g_{ D_s^{\ast} D^{\ast} K^{*} }  \left( \frac{g_{X D^{*} \bar D}}{2}\right) \epsilon_{K^*\ \mu}(k) \ \epsilon_{X}^{\beta} (P-k)  \nonumber \\ 
    & \times
    \int \frac{d^4q}{(2\pi)^4} 
\frac{(P-q)^\rho[(2P-2q-k )^\mu g^{\sigma\alpha} + (2k-P+q )^\sigma g^{\mu\alpha} - (P-q+k )^\alpha g^{\mu\sigma}  ]}{\left[ q^2- m_{\bar D}^2 \right] \left[ (P-q)^2-m_{D_s^*}^2 \right] \left[ (P-q-k)^2- m_{D^*}^2 \right] }
\nonumber \\ 
    & \times 
\left[ -g_{\rho \sigma } + \frac{(P-q)_\rho (P-q)_\sigma}{m_{ D_s^*}^2}\right] \left[ -g_{\alpha \beta } + \frac{(P-q-k)_\alpha (P-q-k)_\beta}{m_{\bar D^*}^2}\right] , 
\nonumber \\     
\mathcal{M}_c &= - i \ g_{ B D_s^{\ast} \bar D^{\ast}}  g_{ D_s^{\ast} D K^{*} }  \left( \frac{g_{X D^{*} \bar D}}{2}\right) \epsilon_{K^*\ \mu}(k) \ \epsilon_{X}^{\beta} (P-k) \nonumber \\ 
    & \times
    \int \frac{d^4q}{(2\pi)^4} 
\frac{q_\chi (P-q)_\nu (P-q)_\gamma k_\delta}{\left[ q^2- m_{\bar D^*}^2 \right] \left[ (P-q)^2-m_{D_s^*}^2 \right] \left[ (P-q-k)^2- m_{D}^2 \right] }
\varepsilon^{\chi \alpha \nu \rho } \varepsilon^{\gamma \sigma \delta \mu }\nonumber \\ 
    & \times 
 \left[ -g_{\alpha \beta } + \frac{q_\alpha q_\beta}{m_{\bar D^*}^2}\right] \left[ -g_{\rho \sigma } + \frac{(P-q)_\rho (P-q)_\sigma}{m_{ D_s^*}^2}\right] , 
 \nonumber \\     
 \mathcal{M}_d  &=  - i \ g_{ B D_s \bar D}  g_{ D_s D^{\ast} K^{*} }  \left( \frac{g_{X D^{*} \bar D}}{2}\right) \epsilon_{K^*\ \mu}(k) \ \epsilon_{X}^{\beta} (P-k) 
 \nonumber \\ 
    &  
    \times
    \int \frac{d^4q}{(2\pi)^4} 
\frac{(P-q-k)_\alpha \ k_\rho \ \varepsilon^{\alpha \gamma \rho \mu }}{\left[ q^2- m_{\bar D}^2 \right] \left[ (P-q)^2-m_{D_s}^2 \right] \left[ (P-q-k)^2- m_{D^*}^2 \right] }
% \nonumber \\ 
%    & \times 
 \left[ -g_{ \gamma \beta  } + \frac{(P-q-k)_\gamma (P-q-k)_\beta}{m_{\bar D^*}^2}\right], 
\label{amplMaMbMc}
\end{align}

The amplitudes $\mathcal{M}_i$ in Eq.~(\ref{amplMaMbMc}) are evaluated according to the following strategy (the complete derivation is lengthy and has been omitted for conciseness): (i) use of the transversality conditions $\epsilon_{K^*}(k) \cdot k = 0$ and  $\epsilon_{X} (P-k) \cdot (P-k) = 0$; (ii) choice of the $K^{\ast 0}  X(3872) $ C.M. frame, which engenders $\vec{P}=0$; (iii) selection of just the positive energy part of the meson propagators, given the fact that the intermediate mesons are heavy particles that propagate mostly with positive energy; (iv) analytical integration over $q_0$, which picks up in the contour below the real axis the pole $q_0 = w_{\bar D^{(*)}} (\vec{q})$, where $w_{\bar D^{(*)}} (\vec{q}) = \sqrt{m_{\bar D^{(*)}}^2  + \vec{q}^2} $~\cite{Sakai:2020ucu,Brandao:2023vyg}; (v) treatment of the integrals with the Passarino-Veltman reduction method, noticing that in the present case the integrals in Eq.~(\ref{amplMaMbMc}) become tensors of rank $1$ or 2 only in terms of the momentum $k$, since the C.M. frame has been adopted~\cite{MartinezTorres:2014son}; (vi) integral over the three-momentum $\vec{q}$ performed with the presence of a Gaussian form factor to prevent ultraviolet divergence:  
\begin{align}
 \int   d^3 q \to 2 \pi \int _{-1} ^{+1} d(\cos{\theta}) \int d|\vec{q}| \ |\vec{q}|^2  \exp{\left(-\frac{2 |\vec{q}|^2}{\Lambda^2}\right)}, 
    \label{FF}
\end{align}
where $\Lambda$ is a free parameter. As usual in other works, it is chosen $\Lambda = 1 \ \mathrm{GeV}$~\cite{Liu:2024ziu,Wu:2023rrp}. 

Finally, with all pieces discussed above, the partial decay width for the $B^0 \to K^{\ast 0}  X(3872)$ decay can be evaluated with the expression 
\begin{eqnarray}
 \Gamma (B^0 \to K^{\ast 0}  X(3872)) = \frac{1}{8\pi}\frac{|\vec{k}|}{ m^2_{B}} \sum_{Pol} |  \mathcal{M} |^2,
  \label{partialdecaywidth}
\end{eqnarray}
where  $|\vec{k}| = \lambda ^{1/2} \left(m^2_{B},m^2_{K^{\ast} },m^2_{X} \right) / (2 m_{B})$ is the magnitude of the three-momentum of the $K^{\ast0}$ meson in the rest frame of $B^0$. As done in Eq.~(\ref{couplBDSD}), the branching ratio $\mathcal{B}$ is then obtained using $ \Gamma (B^0 \to K^{\ast 0}  X(3872)) / \Gamma_ {B}$.

%%%%%%%%%%%%%%%%%%%%%%%%%%%%%%%%%%%%%%%%%%%%%%%%%%%%%%%%%%%%%%%%%%%%
%%%%%%%%%%%%%%%%%%%%%%%%%%%%%%%%%%%%%%%%%%%%%%%%%%%%%%%%%%%%%%%%%%%%
%%%%%%%%%%%%%%%%%%%%%%%%%%%%%%%%%%%%%%%%%%%%%%%%%%%%%%%%%%%%%%%%%%%%
\section{Results}
\label{Results}
%%%%%%%%%%%%%%%%%%%%%%%%%%%%%%%%%%%%%%%%%%%%%%%%%%%%%%%%%%%%%%%%%%%%
%%%%%%%%%%%%%%%%%%%%%%%%%%%%%%%%%%%%%%%%%%%%%%%%%%%%%%%%%%%%%%%%%%%%
%%%%%%%%%%%%%%%%%%%%%%%%%%%%%%%%%%%%%%%%%%%%%%%%%%%%%%%%%%%%%%%%%%%%

\begin{figure}[!htbp]
	\centering
\includegraphics[{width=8.0cm}]{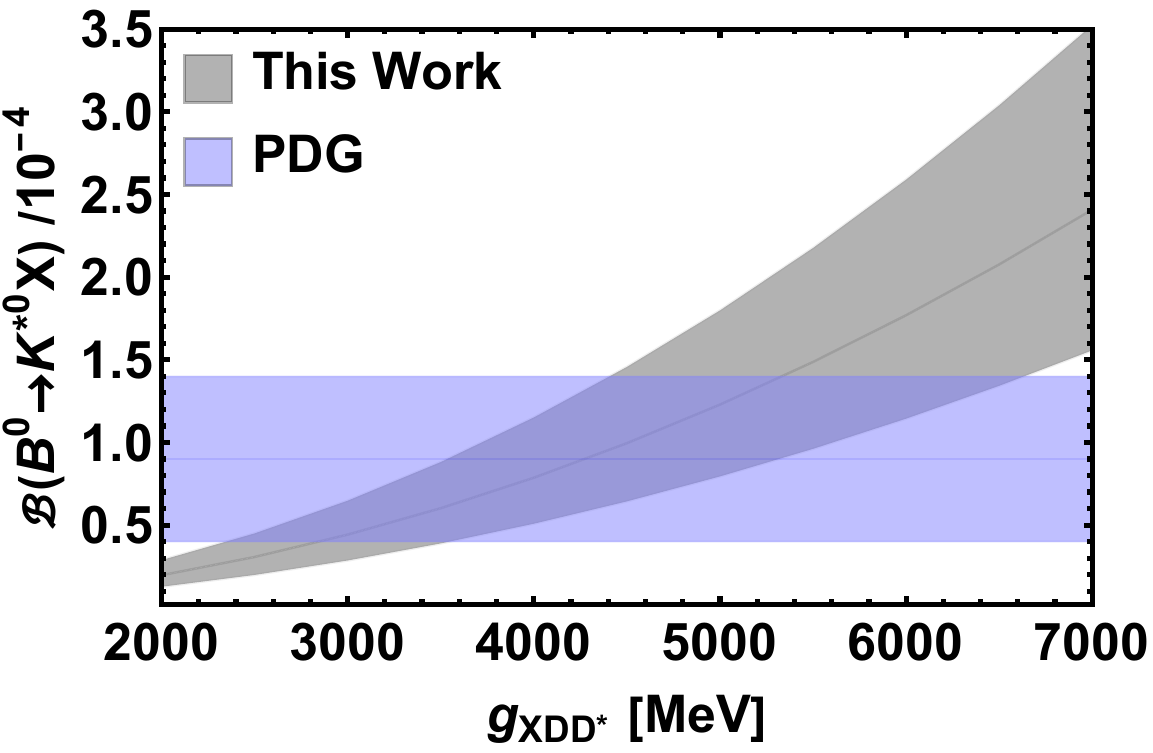} \\
\caption{ Branching ratio of the decay $  (B^0 \to K^{\ast 0}  X(3872)) $ as a function coupling constant $g_{X D^{*} \bar D}$. The gray band denotes the uncertainties coming from the values of the size parameter $ \Lambda$ in the range $0.9\Lambda - 1.1 \Lambda$. The lighter blue band denotes the  available experimental data from Ref.~\cite{ParticleDataGroup:2024cfk}, also reproduced in Table~\ref{tableBR}.}
\label{fig-BR}
\end{figure}

The masses and quantum numbers of the mesons used in the calculations have been taken from from Ref.~\cite{ParticleDataGroup:2024cfk}. 
In Fig.~\ref{fig-BR} the outcomes for the branching ratio of the decay $  (B^0 \to K^{\ast 0}  X(3872)) $ are presented and compared with the available experimental data (from Table~\ref{tableBR}). As pointed in Sec.~\ref{subsec-XDDStar}, no prior preference is assumed for the coupling between the $ X(3872)$ and its charged constituents. In that regard, the findings are exhibited as a function of $ g_{X D^{*} \bar D}$. 
Furthermore, the analysis is also subject to uncertainties originating from the coupling constants $g_{ B D_s^{(\ast)} \bar D^{(\ast)} }$ and $g_{ D_s^{(*)} D^{(*)} K^{*} }$ listed in Table~\ref{tablecouplings}. Propagating these uncertainties yields a combined theoretical error of approximately $40\%$ on the calculated partial decay width. To implement this uncertainty in a straightforward way across the calculation, the form-factor size parameter $\Lambda$ is varied by $\pm 10\%$. This variation produces a comparable overall uncertainty in the final results. Therefore, the predictions are presented as bands corresponding to the range $0.9\Lambda$--$1.1\Lambda$.

It can be seen, as obviously expected, that the branching ratio predictions are sensitive to the coupling, acquiring a bigger magnitude with the increasing $ g_{X D^{*} \bar D}$. Besides,  the variation in the parameter $\Lambda$ within the considered range influences on $\mathcal{B}$.  But most importantly, for $  g_{X D^{*} \bar D} \sim 2400 - 6700 \ \mathrm{MeV}  $ the findings are in a fair agreement with the experimental data, when both theoretical and experimental uncertainties are considered. In particular,  in Table~\ref{Table-BR-coupl} the estimations of $\mathcal{B}$ are shown explicitly for some specific central values of the coupling $ g_{X D^{*} \bar D}$ reported in literature and cited in Sec.~\ref{subsec-XDDStar}. The predictions obtained with the coupling calculated in Refs.~\cite{Gamermann:2009fv,Gamermann:2009uq,Wu:2023rrp,Aceti:2012cb} align well with the experimental measurements.

%%%%%%%%%%%%%%%%%%%%%%%%%%%%%%%%%%%%%%%%%%%%%%%%%%%%%
%%%%%%%%%%%%%%%%%%%%%%%%%%%%%%%%%%%%%%%%%%%%%%%%%%%%%

%%%%%%%%%%%%%%%%%%%%%%%%%%%%%%%%%%%%%%%%%%%%%%%%%%%%%
\begin{table}[!htbp]
\centering
\caption{Predictions of the branching ratio of the decay $  (B^0 \to K^{\ast 0}  X(3872)) $ and the ratio $R$ defined in Eq.~(\ref{ratio}) for some specific central values of $ g_{X D^{*} \bar D}$ reported in literature. The available experimental data are given in Table~\ref{tableBR}).}
\begin{tabular}{ccc}
\hline
\hline
$g_{X D^{*} \bar D}$ [MeV]  & $\mathcal{B} \ (/10^{-5})$ & $R  \ (/10^{-2})$  \\ \hline
 1813~\cite{Dong:2009yp}  &  $1.1-2.4$   &  $0.2-0.5$\\
 2982~\cite{Gamermann:2009fv}  &  $2.8-6.4$   &  $0.6-1.3$\\
 3238~\cite{Gamermann:2009uq}  &  $3.3-7.5$   &  $0.7-1.6$\\
 3390~\cite{Wu:2023rrp}  &  $3.7-8.3$    &  $0.8-1.7$   \\ 
 3638~\cite{Aceti:2012cb}  &  $4.2-9.5$   &  $0.9-2.0$ \\ 
 \hline\hline
\end{tabular}
\label{Table-BR-coupl}
\end{table}

%%%%%%%%%%%%%%%%%%%%%%%%%%%%%%%%%%%%%%%%%%%%%%%%%%%%%

\begin{figure}[!htbp]
	\centering
\includegraphics[{width=8.0cm}]{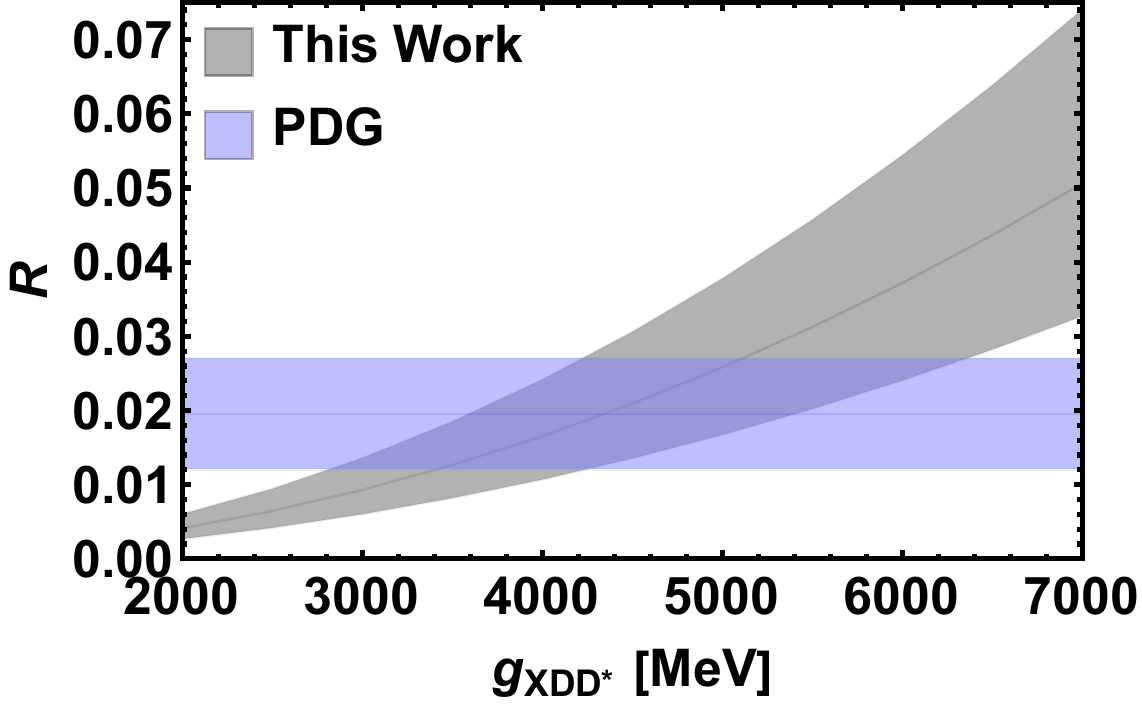} \\
\caption{ Ratio $R$ defined in Eq.~(\ref{ratio}) as a function of $ g_{X D^{*} \bar D}$ as a function coupling constant $g_{X D^{*} \bar D}$. The gray band denotes the uncertainties coming from the values of the size parameter $ \Lambda$ in the range $0.9\Lambda - 1.1 \Lambda$. The lighter blue band denotes the  available experimental data from Ref.~\cite{LHCb:2025vjj}.}
\label{fig-Ratio}
\end{figure}

Furthermore, as pointed in Sec.~\ref{sec-intr}, the LHCb Collaboration reported in Fig.~2 of Ref.~\cite{LHCb:2025vjj} the relative branching ratio $R$ between the $ \mathcal{B} (B^0 \to K^{\ast 0}  X(3872)) \times \mathcal{B}( X(3872) \to J/\psi \pi^{+} \pi^{-} )$ and $ \mathcal{B}(B^0 \to K^{\ast 0}  \psi(2S) )  \times \mathcal{B}( \psi (2S) \to J/\psi \pi^{+} \pi^{-} )$, which is given in Eq.~(\ref{ratio}). With the experimental data of $\mathcal{B}( B^0 \to K^{\ast 0} \psi (2S) )$, $\mathcal{B}( X(3872) \to J/\psi \pi^{+} \pi^{-} )$ and $\mathcal{B}( \psi (2S) \to J/\psi \pi^{+} \pi^{-} )$, given in Table~\ref{tableBR}, this quantity can be also estimated.  Thus, in Fig.~\ref{fig-Ratio} is presented the ratio $R$ defined in Eq.~(\ref{ratio}) as a function of $ g_{X D^{*} \bar D}$, and compared with the available experimental data. 
As in the case of the absolute value of the  branching fraction discussed above, $R$ is influenced by variations of the parameter $\Lambda$, but in the range of $  g_{X D^{*} \bar D} \sim 2800 - 6600 \ \mathrm{MeV}  $ the results are consistent with the experimental data within theoretical and experimental uncertainties. In Table~\ref{Table-BR-coupl} the results based on $g_{X D^{*} \bar{D}}$ from Refs.~\cite{Gamermann:2009fv,Gamermann:2009uq,Wu:2023rrp,Aceti:2012cb} are shown, and yield predictions aligned with experiment.

%%%%%%%%%%%%%%%%%%%%%%%%%%%%%%%%%%%%%%%%%%%%%%%%%%%%%
%%%%%%%%%%%%%%%%%%%%%%%%%%%%%%%%%%%%%%%%%%%%%%%%%%%%%

In the end, the ability of the $\bar{D}^*D$ picture to reproduce the branching ratio $\mathcal{B}( B^0 \to K^{\ast 0}  X(3872) )$ and the ratio $R$ adds to other successful studies of different $X(3872)$ branching fractions in $B$ decays that argue for its dominant molecular nature. 
But a final point deserving emphasis is the role of the charged $( D^{*+}D^- -c.c.)$ component in certain production processes, such as those mediated by the triangle mechanisms depicted in Fig.~\ref{DIAG2} (cf. Ref.~\cite{MartinezTorres:2014son}).
The significant difference in binding energy—approximately 0.2 MeV for the neutral $(\bar{D}^0 D^{*0}-c.c.)$ channel versus 8 MeV for the charged $( D^{*+}D^- -c.c.)$ channel— has a direct consequence on the spatial structure of the wave function. The spatial extent of a bound state is inversely proportional to the square root of its binding energy; therefore the wave function of the loosely bound neutral  component extends far into the asymptotic region, while the wave function of the tightly bound charged component is comparatively compact and localized. This disparity in spatial distribution leads to a much higher probability of finding the system in the  configuration when integrated over all space (see also~\cite{Song:2023pdq}). This is the fundamental reason why one might intuitively expect the charged components to play a minor role in the overall properties of the $X(3872)$. 
However, Ref.~\cite{Gamermann:2009uq} provides the link between the internal structure of a molecular state and its observable decay properties. It was shown that for an $S$-wave bound state formed from two hadrons, the value of its wave function at the origin, $|\psi(0)|$, is directly proportional to the effective coupling constant $g$ between the molecular state and its constituent hadrons ($|\psi(0)|^2 \propto g^2$),
where a larger coupling means a higher probability amplitude for finding the constituents at zero separation. This idea is pivotal for understanding the $X(3872)$. Analyses in Refs.~\cite{Gamermann:2009fv,Gamermann:2009uq,Wu:2023rrp} have determined that the couplings to the neutral and charged channels are comparable in magnitude. 
Consequently, it follows that the squared wave functions at the origin, $|\psi_{\bar{D}^0 D^{*0}}(0)|^2$ and $|\psi_{D^{*+}D^- }(0)|^2$, are also of similar magnitude.  
This has a direct implication for short-range processes, such as strong decays or production mechanisms. These processes are sensitive not to the long-range tail of the wave function (which is dominated by the loosely bound neutral component) but to its value at the origin, where the overlap and interaction occur. Therefore, since both the neutral and charged components contribute significantly to $|\psi(0)|^2$, an accurate theoretical description of any short-range process must include the contributions from both constituents. 

Hence, the outcomes above provide one more practical example of the relevance and the mandatory inclusion of the charged components for a quantitatively correct description of the $X(3872)$. This is the main finding of the present study.

%%%%%%%%%%%%%%%%%%%%%%%%%%%%%%%%%%%%%%%%%%%%%%%%%%%
%%%%%%%%%%%%%%%%%%%%%%%%%%%%%%%%%%%%%%%%%%%%%%%%%%%
%%%%%%%%%%%%%%%%%%%%%%%%%%%%%%%%%%%%%%%%%%%%%%%%%%%
\section{Concluding remarks}
\label{Conclusions}
%%%%%%%%%%%%%%%%%%%%%%%%%%%%%%%%%%%%%%%%%%%%%%%%%%%
%%%%%%%%%%%%%%%%%%%%%%%%%%%%%%%%%%%%%%%%%%%%%%%%%%%
%%%%%%%%%%%%%%%%%%%%%%%%%%%%%%%%%%%%%%%%%%%%%%%%%%%

This work has been devoted to the analysis of the production of the $X(3872)$ in the decay $B^0 \to K^{\ast 0} X(3872)$ via triangle mechanisms within the molecular picture. Assuming the $X(3872)$ to be a $(D\bar{D}^* - \text{c.c.})$ bound state, the decay proceeds through the sequence $B^0 \to D_s^{(*)+} D^{(*)-}$, $D_s^{(*)+} \to K^{\ast 0} D^{(*)+}$, and final-state rescattering $D^{(*)+} D^{(*)-} \to X(3872)$. Using an effective Lagrangian approach, the branching fraction $\mathcal{B}(B^0 \to K^{\ast 0} X(3872))$ and the ratio $R$ relative to the normalization channel $B^0 \to K^{\ast 0} \psi(2S)$ have been computed, examining their dependence on the coupling $g_{X D^{*} \bar{D}}$ of the charged components $(D^+ D^{*-} + \text{c.c.})$ to the $X(3872)$. The predictions obtained using $g_{X D^{*} \bar{D}}$ from Refs.~\cite{Gamermann:2009fv,Gamermann:2009uq,Wu:2023rrp,Aceti:2012cb} yield consistent descriptions of both $\mathcal{B}(B^0 \to K^{\ast 0} X(3872))$ and $R$. Hence, this analysis stands as another demonstration for the necessity of including the charged components to reach a complete molecular description of $X(3872)$ production, such as in $B^0 \to K^{\ast 0} X(3872)$ processes. 

Finally, some questions warrant further discussion. First, the dependence of the results on the parameter $\Lambda$ and the coupling strengths. This highlights the need for future improvements to the model, particularly through the determination of these quantities via more fundamental and rigorous theoretical approaches, which is a non-trivial task. 
Furthermore, the treatment above exclusively considers the $(D\bar{D}^* - \text{c.c.})$ configuration, while other potentially significant components--such as $(D_s\bar{D}_s^* - \text{c.c.})$~\cite{MartinezTorres:2014son}, $J/\psi \rho$, $J/\psi \omega$~\cite{Dong:2009uf,Dong:2009yp}, or even a compact $c\bar{c}$ core~\cite{Dong:2009uf,Dong:2009yp}--remain unaccounted for. The reasonable agreement with branching fractions achieved through the $\bar{D}^{*}D$ picture, while suggestive, does not exclude the necessity of more studies exploring alternative theoretical approaches incorporating different components.

%%%%%%%%%%%%%%%%%%%%%%%%%%%%%%%%%%%%%%%%%%%%%%%%%%%
%%%%%%%%%%%%%%%%%%%%%%%%%%%%%%%%%%%%%%%%%%%%%%%%%%%
%%%%%%%%%%%%%%%%%%%%%%%%%%%%%%%%%%%%%%%%%%%%%%%%%%%
\begin{acknowledgements}
%%%%%%%%%%%%%%%%%%%%%%%%%%%%%%%%%%%%%%%%%%%%%%%%%%%
%%%%%%%%%%%%%%%%%%%%%%%%%%%%%%%%%%%%%%%%%%%%%%%%%%%
%%%%%%%%%%%%%%%%%%%%%%%%%%%%%%%%%%%%%%%%%%%%%%%%%%%

This work was partly supported by the Brazilian CNPq ( Grants No. 400215/2022-5, 308299/2023-0, 402942/2024-8) and CNPq/FAPERJ under the Project INCT-F\'{\i}sica Nuclear e Aplica\c c\~oes (Contract No. 464898/2014-5).

\end{acknowledgements}

%%%%%%%%%%%%%%%%%%%%%%%%%%%%%%%%%%%%%%%%%%%%%%%%%%%%%%%%%%%%%%%%%%%%
%%%%%%%%%%%%%%%%%%%%%%%%%%%%%%%%%%%%%%%%%%%%%%%%%%%%%%%%%%%%%%%%%%%%
\appendix
%%%%%%%%%%%%%%%%%%%%%%%%%%%%%%%%%%%%%%%%%%%%%%%%%%%%%%%%%%%%%%%%%%%%
%%%%%%%%%%%%%%%%%%%%%%%%%%%%%%%%%%%%%%%%%%%%%%%%%%%%%%%%%%%%%%%%%%%%

%%%%%%%%%%%%%%%%%%%%%%%%%%%%%%%%%%%%%%%%%%%%%%%%%%%%%%%%%%%%%%%%%%%%
%%%%%%%%%%%%%%%%%%%%%%%%%%%%%%%%%%%%%%%%%%%%%%%%%%%%%%%%%%%%%%%%%%%%
\section{Effective Framework for the Weak Decays $B^0 \to D_s^{(*)+} D^{(*)-}$}
\label{sec:appa}
%%%%%%%%%%%%%%%%%%%%%%%%%%%%%%%%%%%%%%%%%%%%%%%%%%%%%%%%%%%%%%%%%%%%
%%%%%%%%%%%%%%%%%%%%%%%%%%%%%%%%%%%%%%%%%%%%%%%%%%%%%%%%%%%%%%%%%%%%

%In the analysis of the reaction $B^0 \to K^{\ast 0} X(3872)$, it is advantageous to separate the initial weak decay mechanism from the subsequent low-energy hadronic dynamics. 
This Appendix summarizes the effective method adopted for the $B^0 \to D_s^{(*)+} D^{(*)-}$ primary vertices, which encapsulates the weak decay dynamics into a single vertex structure with an effective coupling constant. 
The standard factorization approach~\cite{Ali:1998eb,Wu:2023rrp} serves as both the starting point and motivation for this simplified methodology.

In essence, the factorization method provides a more fundamental estimate of the production amplitude by separating the weak operator into hadronic matrix elements with momentum-dependent form factors and incorporating the relevant CKM factors and Wilson coefficients. 
Specifically, the fundamental decays $B^0 \to D_s^{(*)+} D^{(*)-}$ proceed via the Cabibbo-favored transition $b \to c\bar{c}s$. In the ``naive" factorization approach~\cite{Ali:1998eb,Wu:2023rrp}, the decay amplitude for $B^0 \to D_s^{(*)+} D^{(*)-}$ factorizes as:
\begin{equation}
\mathcal{M}(B^0 \to D_s^{(*)+} D^{(*)-}) = \frac{G_F}{\sqrt{2}} V_{cb}V_{cs}^* a_1 \langle D^{(*)-} | (\bar{c}b) | B^0 \rangle \langle D_s^{(*)+} | (\bar{s}c) | 0 \rangle + \mathcal{O}(1/N_c),
\label{eq:fact_amplitude_general}
\end{equation}
where $G_F$ is the Fermi constant, $V_{cb}$, $V_{cs}$ are CKM matrix elements, $a_1$ is the corresponding effective Wilson coefficient, $(\bar{q}_1 q_2)$ is the corresponding current, and $N_c$ is the number of colors.

The hadronic matrix elements are parameterized in terms of decay constants and form factors as follows:
    \begin{align}
    \langle D_s^+ | (\bar{s}  c)_{V-A} | 0 \rangle  ^\mu &= i f_{D_s} p_{D_s}^\mu, \nonumber \\
    \langle D_s^{*+} | (\bar{s}\gamma_{\mu}  c ) | 0 \rangle  &= m_{D_s^*} f_{D_s^*} \epsilon_\mu^*, \nonumber \\
    \langle D^- | (\bar{c} b)_{V-A} | B^0 \rangle _\mu &= \left[ (p_B + p_D)_\mu - \frac{m_B^2 - m_D^2}{q^2} q_\mu \right] F_1^{B\to D}(q^2) + \frac{m_B^2 - m_D^2}{q^2} q_\mu F_0^{B\to D}(q^2), \nonumber \\
    \langle D^{*-} | (\bar{c}b)_{V-A} | B^0 \rangle_\mu &= \frac{2i}{m_B + m_{D^*}} \epsilon_{\mu\nu\rho\sigma} \epsilon^{*\nu} p_B^\rho p_{D^*}^\sigma V^{B\to D^*}(q^2) + \cdots , 
    \label{matrix_el1}
    \end{align}
where $f_{D_s^{(*)}}$ is the $D_s^{(*)}$ decay constant; $F_0^{B\to D}(q^2)$,  $F_1^{B\to D}(q^2)$ and $V^{B\to D^*}(q^2)$ are the $B \to D^{(*)}$ transition form factors, $(\bar{q}_1 q_2)_{V-A} \equiv \bar{q}_1 \gamma_\mu (1-\gamma_5) q_2$ is the corresponding $V - A$ current, and $q^\mu = p_B^\mu - p_D^{(*)\mu}$ is the momentum transfer.

The $B \to D^{(*)}$ form factors exhibit momentum dependence typically parametrized as~\cite{Ali:1998eb,Wu:2023rrp}:
\begin{equation}
F(q^2) = \frac{F(0)}{1 - a(q^2/m_B^2) + b(q^2/m_B^2)^2},
\label{eq:general_ff}
\end{equation}
with parameters $a$, $b$ to be determined.

However, the present approach employs a simplified version of the amplitudes for the weak vertices $B^0 \to D_s^{(*)+} D^{(*)-}$ in Eq.~(\ref{eq:fact_amplitude_general}). First, it should be noted that for each specific channel, the momentum transfer renders an unexceptional $q^2$-dependence of the form factors of type (\ref{eq:general_ff}). Specifically, the dominant form factors exhibit moderate variations across the physical $q^2$ range~\cite{Na:2015kha}. These variations are typically smaller than the model dependencies introduced by the hadronic interaction potentials, regularization schemes, and other inputs.

Therefore, given the exploratory nature of the current study, the momentum dependence of the form factors at the weak vertex is omitted, and they are evaluated at the physical $q^2 = m_{D_s^{(*)}}^2$ point. Momentum dependence is incorporated solely in the hadronic rescattering processes. This dependence is regulated by the phenomenological form factor in Eq.~(\ref{FF}), which also prevents ultraviolet divergences. The associated theoretical uncertainties are taken into account by varying the form factor size parameter $\Lambda$ within the range $0.9\Lambda$--$1.1\Lambda$, as discussed in Sec.~\ref{Results}.

Hence, the weak decay dynamics encoded in the matrix elements of Eq.~(\ref{eq:general_ff}) are enclosed in a single effective vertex structure, with all parameters of the factorization method accommodated within a single coupling constant. This prescription can be summarized as:
\begin{equation}
 \frac{G_F}{\sqrt{2}} V_{cb}V_{cs}^* a_1 \left[ f_{D_s^{(*)}} F^{B\to D^{(*)}}(m_{D_s^{(*)}}^2) \times \text{(kinematic factors)} \right] + \cdots \to g_{B D_s^{(*)} D^{(*)}} \times \text{(Lorentz structures)}.
\label{eq:coupling_relation_general}
\end{equation}
This approximation provides a simpler framework that captures the essential physics while maintaining direct contact with experimental data. The  amplitudes for the $B^0 \to D_s^{(*)+} D^{(*)-}$ reactions are given in Eq.~(\ref{amplBDDS}), which implements the appropriate vertex structures. The corresponding numerical values for the effective couplings $g_{B D_s^{(*)} D^{(*)}}$ are calculated using Eq.~(\ref{couplBDSD}) and listed in Table~\ref{tablecouplings}.

%%%%%%%%%%%%%%%%%%%%%%%%%%%%%%%%%%%%%%%%%%%%%%%%%%%%%%%%%%%%%%%%%%%%
%%%%%%%%%%%%%%%%%%%%%%%%%%%%%%%%%%%%%%%%%%%%%%%%%%%%%%%%%%%%%%%%%%%%

%%%%%%%%%%%%%%%%%%%%%%%%%%%%%%%%%%%%%%%%%%%%%%%%%%%%%%%%%%%%%%%%%%%%%%%%
%%%%%%%%%%%%%%%%%%%%%%%%%%%%%%%%%%%%%%%%%%%%%%%%%%%%%%%%%%%%%%%%%%%%%%%%
%\References
%\begin{thebibliography}{99}
%%%%%%%%%%%%%%%%%%%%%%%%%%%%%%%%%%%%%%%%%%%%%%%%%%%%%%%%%%%%%%%%%%%%%%%%
%%%%%%%%%%%%%%%%%%%%%%%%%%%%%%%%%%%%%%%%%%%%%%%%%%%%%%%%%%%%%%%%%%%%%%%%

\bibliographystyle{apsrev4-1}
\bibliography{BtoKStarX}

%\end{thebibliography}

\end{document}